\documentclass[10pt,aps,prd,showpacs,footinbib,twocolumn]{revtex4-1}
\usepackage[usenames,dvipsnames]{color}
\usepackage{amsfonts}
\usepackage{amssymb}
\usepackage{amsmath}
\usepackage{graphicx}
\usepackage[colorlinks]{hyperref}
\usepackage[normalem]{ulem}

\begin{document}

\title{Trace of the energy-momentum tensor and macroscopic properties of neutron stars}

\author{Dylan M.\ Podkowka}
\email{dpodkowk@uoguelph.ca}
\affiliation{Department of Physics, University of Guelph, Guelph, Ontario, N1G 2W1, Canada.}

\author{Raissa F.\ P.\ Mendes}
\email{rfpmendes@id.uff.br}
\affiliation{Instituto de F\'isica, Universidade Federal Fluminense, Niter\'oi, Rio de Janeiro, 24210-346, Brazil.}

\author{Eric Poisson}
\email{epoisson@uoguelph.ca}
\affiliation{Department of Physics, University of Guelph, Guelph, Ontario, N1G 2W1, Canada}

\date{\today}

\begin{abstract}
A generic feature of scalar extensions of general relativity is the coupling of the scalar degrees of freedom to the trace $T$ of the energy-momentum tensor of matter fields. Interesting phenomenology arises when the trace becomes positive---when pressure exceeds one third of the energy density---a condition that may be satisfied in the core of neutron stars. In this work, we study how the positiveness of the trace of the energy-momentum tensor correlates with macroscopic properties of neutron stars. We first show that the compactness for which $T=0$ at the stellar center is approximately equation-of-state independent, and given by $C = 0.262_{-0.017}^{+0.011}$ (90\% confidence interval). Next, we exploit Bayesian inference to derive a probability distribution function for the value of $T$ at the stellar center given a putative measurement of the compactness of a neutron star. This investigation is a necessary step in order to use present and future observations of neutron star properties to constrain scalar-tensor theories based on effects that depend on the sign of $T$.
\end{abstract}


\maketitle

\section{Introduction and Summary} \label{sec:intro}

A generic feature of scalar extensions of general relativity (GR) is the coupling of the scalar degrees of freedom to the trace of the energy-momentum tensor of matter fields. Indeed, a typical field equation in scalar-tensor gravity would have the schematic form \cite{Damour1992}
\begin{equation}
\Box_g \phi - \frac{dV(\phi)}{d\phi}= - \alpha(\phi) T,
\end{equation}
where $\Box_g$ denotes the covariant wave operator, defined in terms of a derivative operator compatible with a metric $g_{\mu\nu}$ that obeys some modified version of Einstein's equations, $\alpha(\phi)$ is a coupling function, $V(\phi)$ is a potential term, and $T := g_{\mu\nu} T^{\mu\nu}$, where $T^{\mu \nu} := (2/\sqrt{-g}) \delta S_m /\delta g_{\mu \nu} $, with $S_m$ denoting the action for the matter fields. 

Interestingly, new phenomenology may arise in scalar-tensor theories when $T$ changes sign. For instance, it was shown in Refs.~\cite{Mendes2015,Palenzuela2016,Mendes2016} that scalar-tensor theories that reproduce the predictions of GR in the regime of weak gravitational fields \cite{Damour1996a,Anderson2017} may deviate considerably from GR around neutron stars (NSs) when $T$ becomes positive in the stellar interior. The new effects, which include spontaneous scalarization \cite{Damour1993} of the star or gravitational collapse to a black hole, could leave observable signatures in electromagnetic and gravitational-wave data, and enable unique tests of these theories. Instabilities that depend on the positiveness of $T$ were also identified in theories with screening mechanisms \cite{Babichev2010,Brax2017}. Importantly, these effects depend on the value of $T$ inside general-relativistic stars, and not inside equilibrium configurations already altered by the presence of the scalar field.

In order to explore the full potential of these effects in constraining scalar-tensor theories of gravity, a necessary step is to understand how the positiveness of $T$ is connected to macroscopic, observable properties of neutron stars. This is the question addressed in this paper.

For a perfect fluid with energy density $\epsilon$ and pressure $p$ in the fluid frame, and 4-velocity $u^\mu$ of fluid elements, the energy-momentum tensor is given by 
\begin{equation}
T^{\mu\nu} = (\epsilon + p) u^\mu u^\nu + p g^{\mu\nu},
\end{equation}
and its trace is $T = 3p - \epsilon$. Because the condition $3p\leq \epsilon$ holds for the electromagnetic field and for a system of non-interacting particles, it is sometimes assumed to hold in all generality \cite{Landau1951}. However, as was pointed out long ago by Zel'dovich, relativistic invariant, causal theories describing strongly interacting systems may display the property $3p>\epsilon$ \cite{Zeldovich1962}. This condition does not imply violation of causality, which is embodied in the requirement that the sound speed $c_s \equiv  \sqrt{\partial p/\partial \epsilon}$ be subluminal, or of any energy condition. 
Indeed, many theoretical equations of state (EoS) for neutron stars predict that $T$ should become positive in the core of the most massive and compact configurations \cite{Haensel2007}. 

In fact, to date much uncertainty remains regarding the EoS for cold dense matter well above the nuclear saturation density, given by $n_{\rm ns} \approx 0.16~{\rm fm}^{-3}$ in terms of the baryon number density, or $\rho_{\rm ns} \approx 2.7 \times 10^{14}~{\rm g/cm}^3$ in terms of the baryon mass density. The theoretically proposed models vary considerably in their assumptions on the microscopic constitution of ultradense matter, ranging from pure nucleonic models to models including hyperons, condensates formed of mesons, or quark matter \cite{Haensel2007,Ozel2016}. Not all of these models give rise to stable stars inside of which $T>0$. This feature depends crucially on the stiffness of the EoS, and is less likely to occur for models that include hadronic degrees of freedom which soften the EoS at high densities, such as hyperons or quarks \cite{Mendes2015}. However, it is definitely the case that the current uncertainty on the nuclear EoS leaves ample room for us to entertain the possibility that the condition $T>0$ is satisfied inside some neutron stars in Nature.

\begin{figure}[t]
\includegraphics[width=0.48\textwidth]{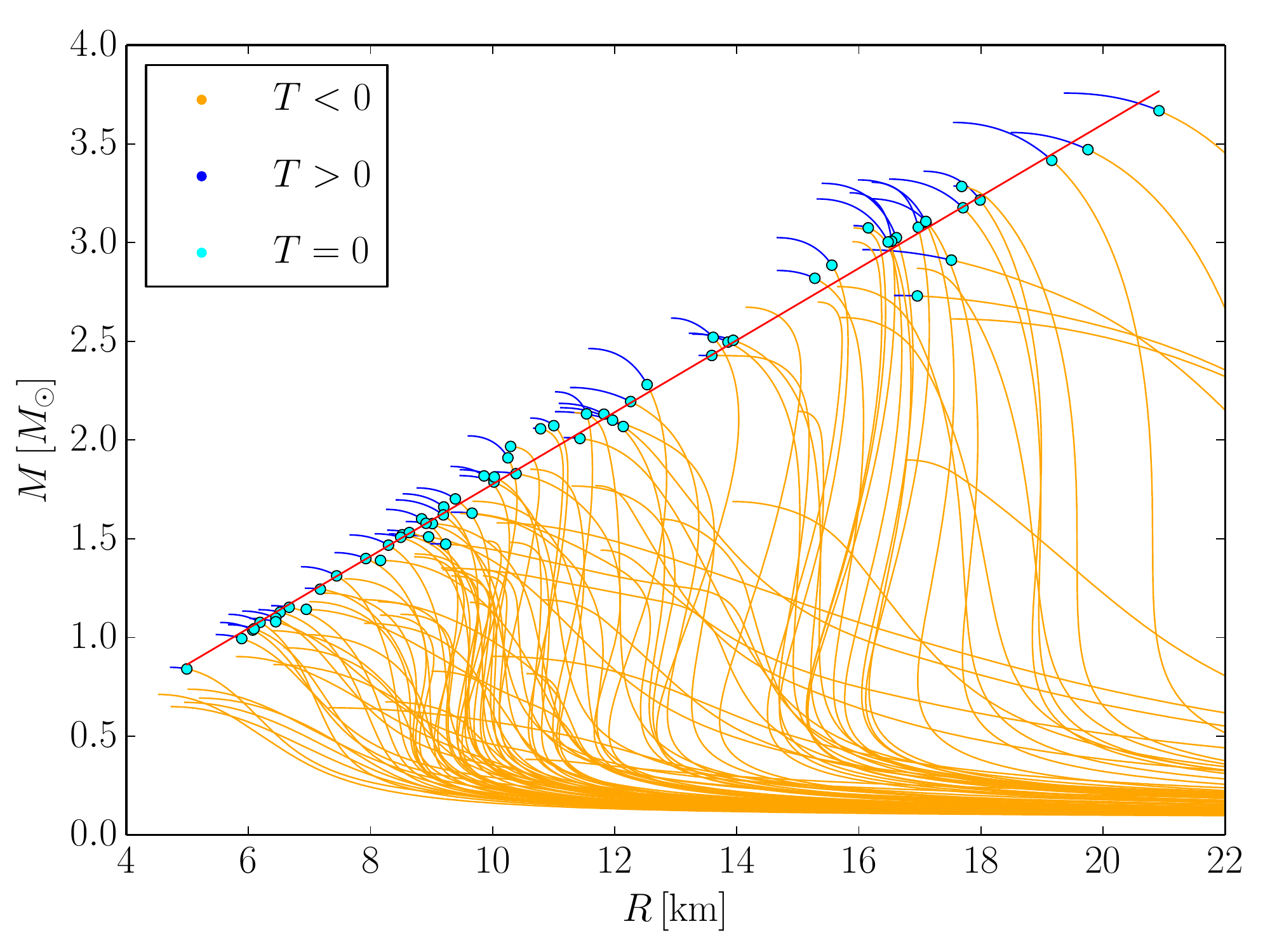}
\caption{Mass-radius curves for 136 piecewise-polytropic causal EoS (cf.~Sec.~\ref{sec:eos}). Configurations for which $T<0$ in the entire stellar interior are displayed in orange, while those for which $T>0$ in some region inside the star are depicted in blue. We highlight in cyan the configurations for which $T=0$ at the stellar center, which marks the transition between the previous regimes. The red dashed line corresponds to $M = (c^2/G) C R$, with  $C = 0.262$, as discussed in the main text.}
\label{fig:massradius}
\end{figure}

To add concreteness to the discussion, in Fig.~\ref{fig:massradius} we display mass-radius curves for several equations of state, drawn from a space of phenomenologically parametrized models described in Sec.~\ref{sec:eos}. Solutions for which $T>0$ in a region of the stellar interior are displayed in blue. Notice that not all EoS admit such configurations, as was mentioned before. Crucially, Fig.~\ref{fig:massradius} shows that, for different EoS, the point along a sequence of equilibrium solutions at which $T$ first becomes positive does not have a unique mass $M$ or radius $R$, but does have a quasi-universal compactness $C=GM/(Rc^2)$, where $G$ is Newton's constant and $c$ is the speed of light. This was already noticed in Ref.~\cite{Mendes2015} for a few EoS, but in Sec.~\ref{sec:transition} this property is quantified more precisely, and we determine the critical compactness as $C = 0.262_{-0.017}^{+0.011}$ (90\% confidence interval). 

The stellar compactness thus seems to be the macroscopic property that best relates to the microscopic condition $T>0$. 
The compactness of a neutron star can be directly inferred, at least in principle, from the measurement of the gravitational redshift of spectral lines produced at the surface of the star \cite{Cottam2002}. But the actual measurement is difficult, and subject to systematic errors \cite{Cottam2008}. Alternatively, one can consider joint measurements of neutron star masses and radii. Presently, masses of approximately 40 neutron stars are known precisely, thanks mainly to the observation and timing of pulsars in binary systems; the determination of neutron star radii has proven more elusive. These measurements typically rely on the detection of thermal X-ray emissions from the surface of the star, combined with distance estimates, and they still face large uncertainties \cite{Ozel2016}. Nonetheless, the precise determination of neutron star radii is a major scientific goal for current and future X-ray missions, such as NICER \cite{Gendreau2016} and LOFT \cite{Feroci2012}, as they can provide invaluable information about the nuclear EoS. Complementary information on NS radii may also be provided by the measurement of the moment of inertia of exquisitely timed pulsars \cite{Raithel2016}. Moreover, gravitational-wave measurements from binary neutron star systems, such as GW170817 \cite{Abbott2017a}, can also constrain the nuclear EoS and provide radius estimates, as the waveform carries information about each neutron star's tidal deformability and possibly its oscillation frequencies. Indeed, the gravitational-wave event GW170817 already enabled the LIGO and Virgo collaborations to set upper limits on the tidal deformabilities of the binary components \cite{Abbott2017a} and to estimate their radii \cite{LIGO2018}, favoring EoS that produce more compact stars.

\begin{figure}[b]
\includegraphics[width=0.48\textwidth]{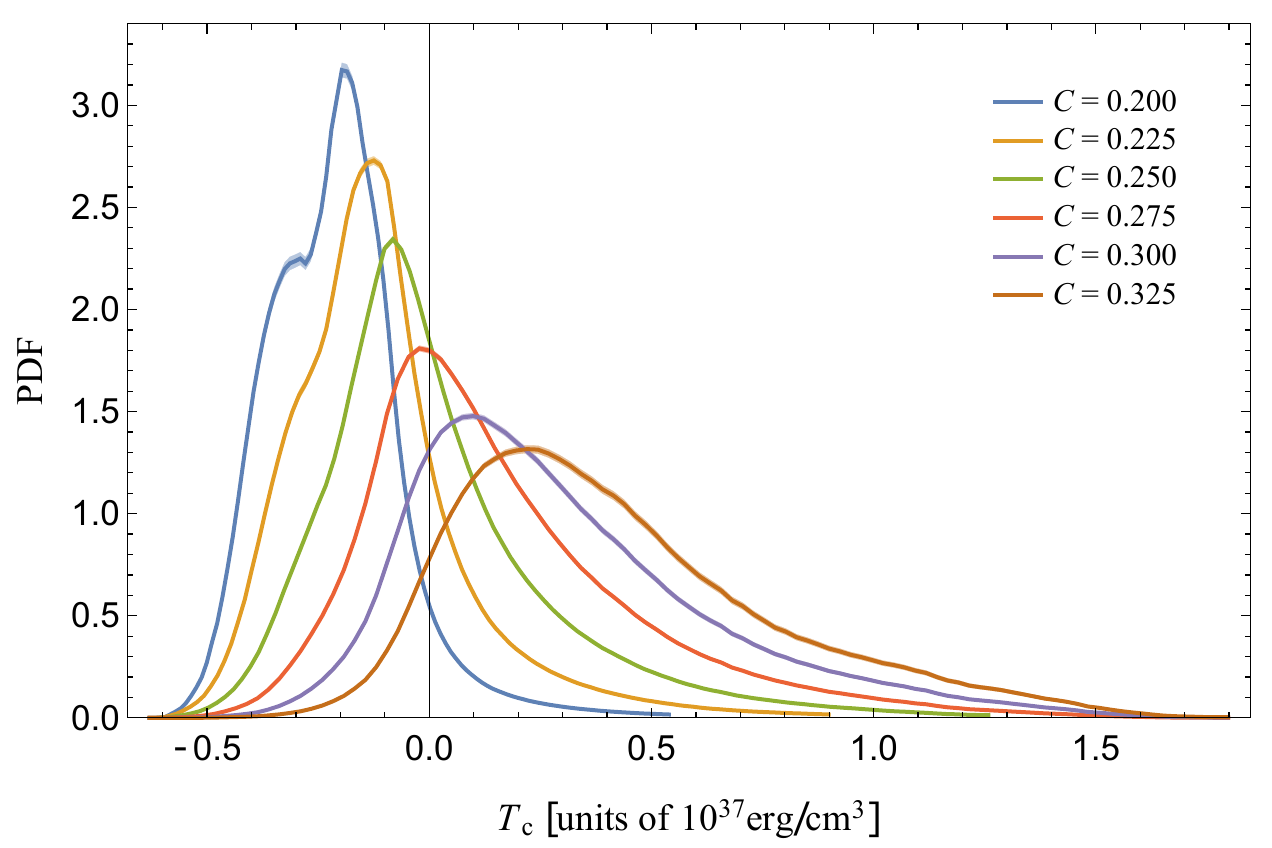}
\caption{Probability distribution functions for $T_c$, the central value of the trace of the energy-momentum tensor, corresponding to  six values of a compactness measurement, ranging from $C=0.200$ (leftmost curve) to $C=0.325$ (rightmost curve). The assumed uncertainty in the measurement is fixed to $\sigma_C = 0.03$ in all cases. Shades give an estimate of the error in the marginalization procedure (cf.~Sec.~\ref{sec:bayesian}).}
\label{fig:sigmafixed}
\end{figure}

Our analysis in Sec.~\ref{sec:transition} shows, in particular, that if the radius of a $2 M_\odot$ NS is measured to be smaller than approximately 10.7 km, then we can ascertain with 90\% confidence that $T>0$ in a region of the stellar interior. Note that such a value for the stellar radius is entirely consistent with current spectroscopic measurements \cite{Ozel2016a}.

Next, we assume that relatively accurate measurements of neutron star properties will be made in future years, and explore, in a Bayesian framework, how these measurements could be translated into a probability distribution function (PDF) for $T_c$, the central value of the trace of the energy-momentum tensor. We describe the procedure and its underlying assumptions in Sec.~\ref{sec:bayesian}. Figure \ref{fig:sigmafixed} shows a sample of our results: we display the PDF for $T_c$ for selected values of a hypothetical measurement of the stellar compactness. If a sufficiently high NS compactness is measured in the coming years, the probability distribution for $T_c$ can be directly translated into constraints on scalar-tensor theories of gravity based on effects such as those described in Refs.~\cite{Mendes2015,Palenzuela2016,Mendes2016}. Further discussion is deferred to Secs.~\ref{sec:bayesian} and \ref{sec:conclusions}.

In the remainder of the paper we adopt geometrized units with $G = c = 1$. 

\section{Setting} \label{sec:setting}

\subsection{Models for the nuclear equation of state} \label{sec:eos}

An equation of state (EoS) is a pair of equations, $p = p(\rho)$ and $\epsilon = \epsilon(\rho)$, which relate the pressure $p$ and energy density $\epsilon$ to the rest-mass density $\rho$ (also known as baryon-mass density). We adopt a phenomenological parametrization for the nuclear EoS consisting of polytropic phases, where
\begin{equation}\label{eq:polytrope}
p(\rho) = K_i \rho^{\Gamma_i}, \quad \rho_{i-1} \leq \rho \leq \rho_i,
\end{equation}
which are joined together continuously at the dividing rest-mass densities $\rho_i$.  The energy density $\epsilon$ is obtained from the first law of thermodynamics for a homentropic fluid, $d(\epsilon/\rho) = - p d(1/\rho)$, which yields
\begin{equation}
\epsilon(\rho) = (1 + a_i) \rho  + \frac{K_i}{\Gamma_i -1} \rho^{\Gamma_i},  \quad \rho_{i-1} \leq \rho \leq \rho_i,
\end{equation}
where $a_i$ is an integration constant given by $a_i = \epsilon(\rho_{i-1})/\rho_{i-1} - 1 - K_i \rho_{i-1}^{\Gamma_i-1}/(\Gamma_i - 1)$. 

A variety of piecewise-polytropic parametrizations for the EoS have been proposed in the literature \cite{Read2009,Steiner2010,Hebeler2013,Steiner2013,Raithel2016a}, and here we adopt the four-parameter model of Read et.~al \cite{Read2009}. (See Ref.~\cite{Raaijmakers2018} for a discussion of possible shortcomings of this parametrization when one seeks to determine EoS parameters from a set of measured NS properties. These are of no concern to us in this paper.) Namely, the EoS at low densities is fixed (to the piecewise-polytropic approximation \cite{Read2009} of the EoS of Ref.~\cite{Douchin2001}) and is matched to a polytrope with adiabatic exponent $\Gamma_1$. At a fixed density $\rho_1 = 10^{14.7}~$g/cm$^3$ ($\rho_1 \approx 1.85 \rho_{\rm ns}$) and pressure $p_1 = p(\rho_1)$, the EoS is joined to a second polytropic phase characterized by the exponent $\Gamma_2$. Finally, at $\rho_2 = 10^{15}~$g/cm$^3$ ($\rho_2 \approx 3.7 \rho_{\rm ns}$), the EoS transitions to a third phase with exponent $\Gamma_3$. The parameters $p_1$ and $\Gamma_1$ essentially determine the overall radius of an equilibrium configuration, while $\Gamma_2$ sets the slope of the mass-radius curve and $\Gamma_3$ roughly determines the maximum mass \cite{Ozel2016a}. The constants $K_i$ ($i=1,2,3$) in Eq.~(\ref{eq:polytrope}) are determined by continuity: $K_{i+1} = p(\rho_i)/\rho_i^{\Gamma_{i+1}}$. 

We restrict the ranges of the free parameters $\{p_1,\Gamma_1,\Gamma_2,\Gamma_3\}$ to $33.5 < \log [p_1/(\textrm{dyne cm}^{-2})] \leq 35.5 $, $1.4 < \Gamma_1 \leq 5$, $1 < \Gamma_2 \leq 5$, and $1 < \Gamma_3 \leq 5$, which were shown to accommodate a diversified set of theoretically proposed EoS \cite{Read2009}. We exclude values of $p_1$ and $\Gamma_1$ which are incompatible, i.e., for which the first polytropic phase falls short of reaching the specified value $p_1$ \cite{Read2009}. 

In the following sections, we will often restrict the range of parameters $\{p_1,\Gamma_1,\Gamma_2,\Gamma_3\}$ further, so that the resulting EoS is causal or complies with basic astrophysical requirements. An EoS will be considered to be causal if the sound speed is subluminal, $c_s := \sqrt{\partial p/\partial \epsilon} < c$, inside all stable configurations. A milder restriction, such as $c_s \lesssim 1.1 c$, is sometimes adopted in the literature on piecewise-polytropic models, with the idea that the transition between phases would be smoother in more realistic EoS, leading to a smaller value of $c_s$ \cite{Read2009}. 
Here we will opt for the more stringent constraint. Additionally, a lower bound will often be imposed on the maximum mass allowed by the EoS. This is necessary to account for the large observed masses of some neutron stars, such as $(2.01 \pm 0.04)M_\odot$ for the pulsar PSR J0348 + 0432 \cite{Antoniadis2013}. We adopt the conservative lower bound $M_{\max} \geq 1.95 M_\odot$. Our precise assumptions on the EoS parameters $\{p_1,\Gamma_1,\Gamma_2,\Gamma_3\}$ will be stated at each point of our analysis.

\subsection{Hydrostatic equilibrium}

The equations governing the hydrostatic equilibrium of a spherically symmetric, static star with line element
\begin{equation}
ds^2 = - e^{2\psi} dt^2 + \frac{1}{1-2m/r} dr^2 + r^2 (d\theta^2 + \sin^2 \theta d\phi^2)
\end{equation}
are given by
\begin{equation}\label{eq:tov}
\frac{dm}{dr} = 4\pi r^2 \epsilon , \quad
\frac{dp}{dr} = - \frac{(\epsilon+p)(m+4\pi r^3 p)}{r^2(1-2m/r)},
\end{equation}
as well as $d\psi/dr = -(\epsilon + p)^{-1} dp/dr$. We assume that an EoS has been specified as in Sec.~\ref{sec:eos}, relating the pressure $p$ and energy density $\epsilon$ to the rest-mass density $\rho$.

Instead of directly integrating Eq.~(\ref{eq:tov}), we shall adopt a modified version of the enthalpy formulation of Ref.~\cite{Lindblom1992a}. The specific enthalpy is defined as
\begin{equation}
h = (\epsilon + p)/\rho,
\end{equation}
and it can replace $\rho$ as the EoS parameter if we determine $\epsilon = \epsilon(h)$ and $p = p(h)$.
The definition of $h$ and the first law imply that $dh/h = dp/(\epsilon + p)$. Therefore, $d\psi = - d\ln h$, and integration returns $e^{2\psi} = (1-2M/R)/h^2$, where we used $h(R) = 1$ and $e^{2\psi(R)} = 1-2M/R$ by continuity with the external Schwarzschild metric. We make the change of variables
\begin{equation}
m = \frac{4\pi}{3} \epsilon_c r^3 \omega, \qquad
h = 1 + (h_c -1) \theta, \quad r^2 = r_0^2 \zeta,
\end{equation}
with $\epsilon_c:= \epsilon (r=0)$, $h_c := h(r=0)$, and $r_0^2 := 3(h_c - 1)/(2\pi \epsilon_c)$. With $\theta$ selected as independent variable, the structure equations (\ref{eq:tov}) become
\begin{align}
\frac{d\zeta}{d\theta} & = - \frac{1-4(h_c-1)\zeta\omega}{[1+ (h_c - 1)\theta](\omega+3p/\epsilon_c)},  \\
\frac{d\omega}{d\theta} &= - \frac{3}{2\zeta} \frac{(\epsilon/\epsilon_c - \omega)[1-4(h_c - 1)\zeta \omega]}{[1+ (h_c - 1)\theta](\omega + 3p/\epsilon_c)}.
\end{align}
These equations are integrated in the interval $1 \geq \theta \geq 0$, with initial conditions $\zeta(\theta = 1) =0$ and $\omega (\theta = 1) = 1$. The surface values $\zeta_s := \zeta(\theta = 0)$ and $\omega_s := \omega (\theta = 0)$ enable a computation of the total mass $M$ and stellar radius $R$. The compactness is given by $M/R = 2 (h_c - 1) \zeta_s \omega_s$. The main advantage of the enthalpy formulation is that the integration limits for the structure equations are known explicitly, instead of determined by a search for the surface. 

\begin{figure}[thb]
\includegraphics[width=0.48\textwidth]{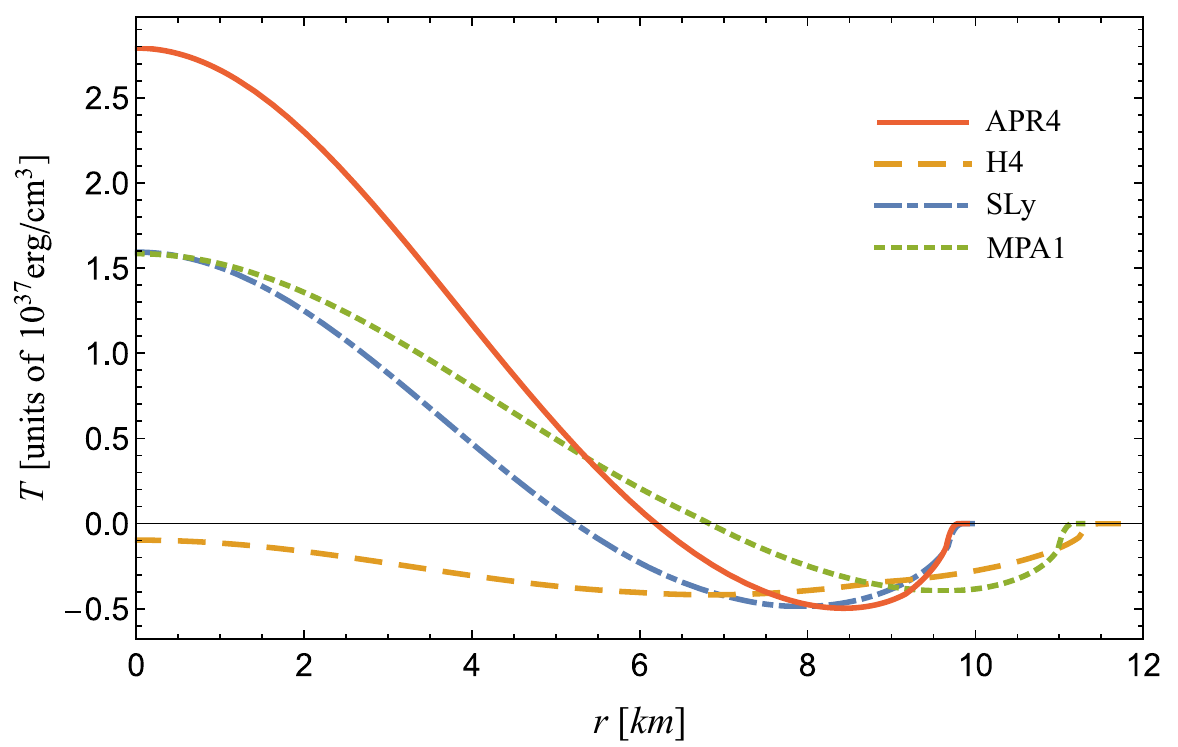}
\caption{$T = 3p - \epsilon$ as a function of the areal radial coordinate for the most massive star allowed by four realistic EoS. We adopt the piecewise-polytropic approximation to the APR4, H4, SLy, and MPA1 models, as given in Table III of Ref.~\cite{Read2009}.}
\label{fig:traceprofile}
\end{figure}

In Fig.~\ref{fig:traceprofile} we plot the radial profile of the trace of the energy-momentum tensor, $T=3p - \epsilon$, for the most massive configuration allowed by four realistic EoS, in their piecewise-polytropic representation given in Ref.~\cite{Read2009}. At the surface of the star, $T$ vanishes, and it is negative near the surface where it is dominated by the rest-mass contribution to the energy density. In the stellar core, $T$ increases as the pressure builds up and, for some EoS, it can become positive in a region of the stellar interior. Note that at asymptotically high densities, $\rho \to \infty$, the theory of quantum chromodynamics predicts the deconfinement of quarks and a free-quark gas behavior, for which again $T \to 0$ \cite{Haensel2007}. For the densities and pressures found in the core of neutron stars, the behavior of matter is still poorly understood, and a transition to $T > 0$ is not ruled out by known nuclear physics. 

\section{Properties of a star with $T_c=0$} \label{sec:transition}

\begin{figure}[b]
\includegraphics[width=0.47\textwidth]{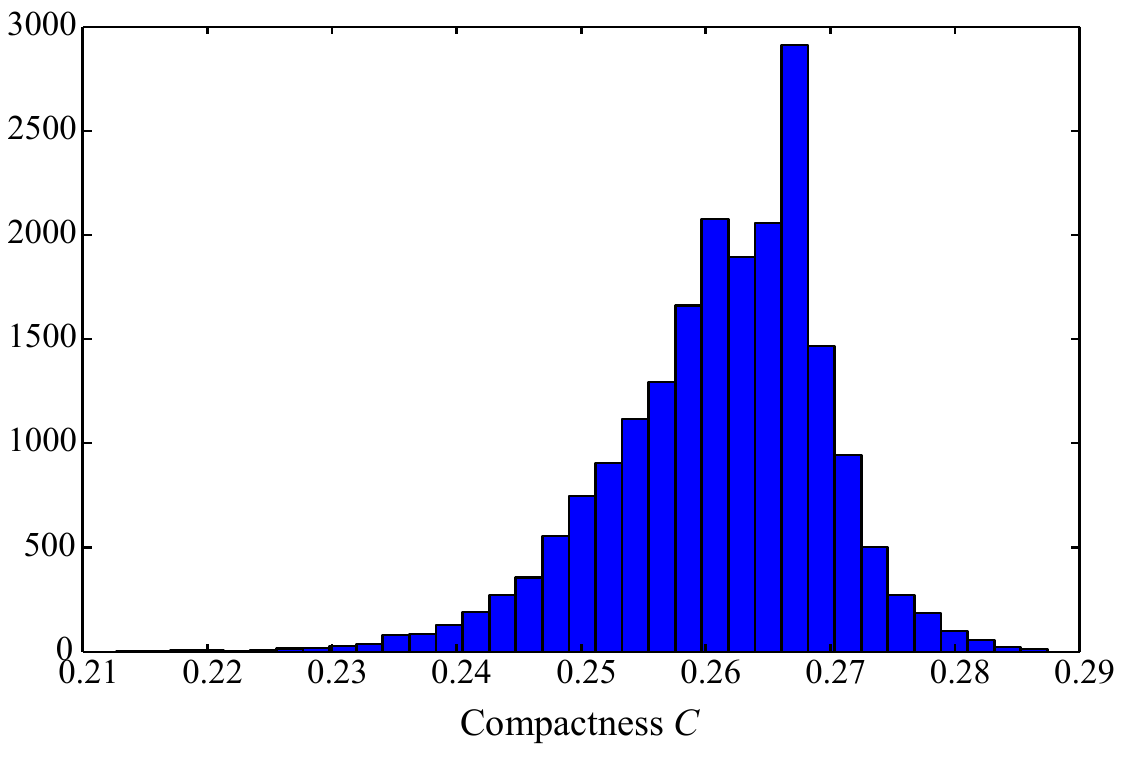}
\caption{Histogram of the compactness of a stable and causal star with $T_c = 0$, computed for 20,000 EoS.}
\label{fig:histogramC}
\end{figure}

As was already anticipated in Sec.~\ref{sec:intro} (cf.~Fig.~\ref{fig:massradius}), for an EoS to allow $T>0$ inside a NS, the basic requirement is that it must support sufficiently compact stable configurations.
In this section we investigate the properties of the critical solution along each equilibrium sequence for which $T_c := T(r=0) = 0$. We sample the EoS parameters uniformly in $\log(p_1)$, $\Gamma_1$, $\Gamma_2$, and $\Gamma_3$, and for each sample we construct the equilibrium configuration with $T_c=0$. If this solution is stable and causal, we store its mass and radius.
As expected from the analysis of Fig.~\ref{fig:massradius}, the mass and radius distributions of solutions with $T_c=0$ carry essentially no information, since they span the entire range of astrophysically plausible masses and radii for NSs. However, the compactness distribution is much sharper, as the histogram in Fig.~\ref{fig:histogramC} shows for a sample of 20,000 EoS. We find that the median and 90\% confidence interval for the compactness of a NS with $T_c =0$ are
\begin{equation}\label{eq:criticalC}
C_{T_c =0} = 0.262_{-0.017}^{+0.011}.
\end{equation}

If one considers a neutron star in the mass range $1.97$--$2.05 M_\odot$, such as the pulsar PSR J0348 + 0432 \cite{Antoniadis2013}, then the compactness distribution in Fig.~\ref{fig:histogramC} can be translated into the critical radius for which the star would have $T_c = 0$. We find $R^{2M_\odot}_{T_c=0} = 11.2^{+1.0}_{-0.5}$ km. Therefore, if this pulsar's radius was measured to be $\lesssim 10.7$ km, then it would be possible to ascertain with 90\% confidence that $T$ is positive in a region of the stellar interior. Such a value for a NS radius is entirely consistent with estimates coming from spectroscopic measurements. In particular, in Ref.~\cite{Ozel2016a} the radius of a $1.5 M_\odot$ NS was estimated to lie in the $10.1$--$11.1$ km range, with previous works on quiescent low-mass X-ray binaries reporting an even smaller typical NS radius of $9.4 \pm 1.2$ km \cite{Guillot2014}. 

It is often reasonable to consider the radius to be approximately constant for NSs in the astrophysically relevant mass range, at least within measurement uncertainties. Therefore, assuming that all NS radii lie in the $10.1$--$11.1$ km range \cite{Ozel2016a}, one can determine how massive a star should be in order that $T = 0$ at the stellar center. The corresponding mass distribution has a median and 90\% credible interval of $1.88^{+0.13}_{-0.16} M_\odot$. Therefore, it is plausible that the most massive observed neutron stars have $T>0$ in their interior.

\begin{figure}[t]
\centering
\includegraphics[width=0.45\textwidth]{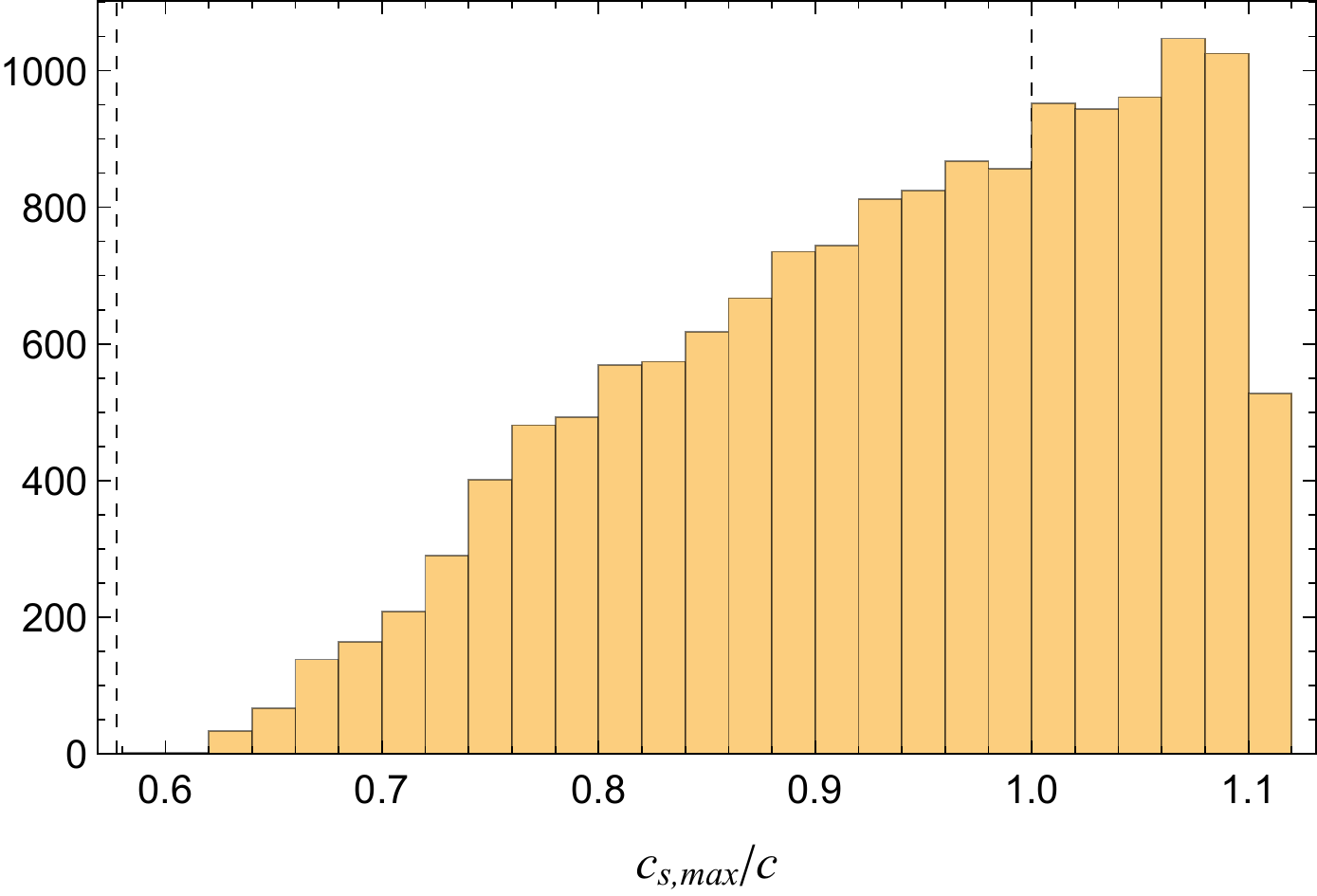}
\caption{Histogram of the maximum speed of sound inside a star with $T_c = 0$, computed for 15,000 sampled EoS. Vertical lines highlight the values $c_{s,{\rm max}} = c/\sqrt{3}$ and  $c_{s,{\rm max}}= c$.}
\label{fig:histogramcs}
\end{figure}

In Fig.~\ref{fig:histogramcs} we display a histogram of the maximum speed of sound $c_{s,{\rm max}}$ inside a star with $T_c = 0$, with uniformly sampled EoS parameters. No further constraints are imposed on the EoS. The histogram is broadly distributed between $c_{s,{\rm max}} \approx 0.60 c$ and $c_{s,{\rm max}} \approx 1.12 c$, with a median of $0.95 c$. We recall that configurations with $c_{s,{\rm max}} > c$ were rejected to construct the histogram of Fig.~\ref{fig:histogramC} and the associated confidence interval.  

\section{Bayesian inference of $T_c$} \label{sec:bayesian}

\begin{figure}[thb]
\includegraphics[width=0.48\textwidth]{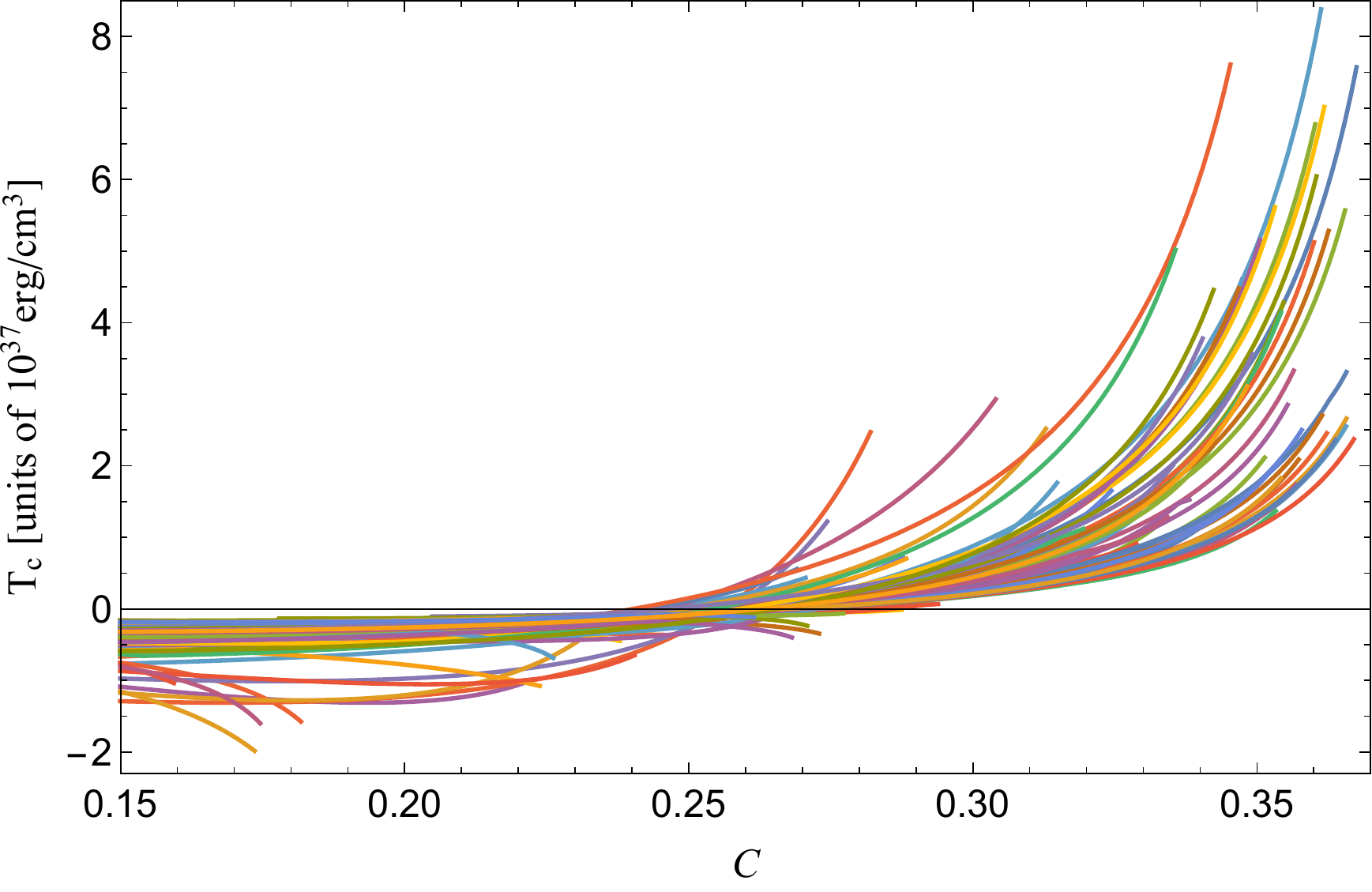}
\caption{Central value of the trace of the energy-momentum tensor $T_c$ as a function of the stellar compactness $C$. Curves are drawn for 100 causal EoS, and they terminate at the maximum value of $C$ for a stable configuration according to each EoS. It should be noted that while $T_c$ is a single-valued function of $C$ for each EoS, $C$ can be a multi-valued function of $T_c$.}
\label{fig:TcvsC}
\end{figure}

The stellar compactness is the macroscopic property of a neutron star that seems to be related to $T_c$ in the most EoS-independent manner. Their relation is illustrated in Fig.~\ref{fig:TcvsC} for 100 piecewise-polytropic EoS, with $\log(p_1)$, $\Gamma_1$, $\Gamma_2$, and $\Gamma_3$ drawn uniformly from the intervals discussed in Sec.~\ref{sec:eos}, and restricted so that the EoS is causal and supports NSs at least as massive as $1.95 M_\odot$.

\subsection{Bayesian framework} 

In this section we determine $p(T_c|\vec{D},I)$, the probability distribution function for $T_c$, the central value of the trace of the energy-momentum tensor, given a set of measured properties $\vec{D}$ for a neutron star and some background information $I$. This can be computed by marginalizing $p(T_c,\vec{\theta}|\vec{D},I)$, the probability that a star of properties $\vec{D}$ has a central trace $T_c$ and EoS parameters $\vec{\theta} = \{\log(p_1),\Gamma_1,\Gamma_2,\Gamma_3\}$,   
\begin{equation} 
p(T_c|\vec{D},I) = \int d\vec{\theta} \, p(T_c, \vec{\theta} | \vec{D}, I).  
\label{eq:main1} 
\end{equation} 
This, in turn, can be obtained from Bayes' theorem, 
\begin{equation} 
p(T_c, \vec{\theta} | \vec{D}, I) = \mathcal{N}\, 
p(\vec{D}|T_c,\vec{\theta},I) p(T_c,\vec{\theta}|I),
\label{eq:main2}
\end{equation}
where $p(\vec{D}|T_c,\vec{\theta},I)$ is the probability that a star with central trace $T_c$ and EoS parameters $\vec{\theta}$ possesses the properties $\vec{D}$, $p(T_c,\vec{\theta}|I)$ is the prior probability on $T_c$ and $\vec{\theta}$, and $\mathcal{N}$ is a normalization constant that can be determined \textit{a posteriori}. 

The marginalization over the EoS parameters in Eq.~(\ref{eq:main1}) is carried out via Monte Carlo integration, with
\begin{equation}\label{eq:montecarlo}
\int d\vec{\theta} \, p(T_c, \vec{\theta} | \vec{D}, I) \approx \frac{V_{\rm EoS}}{N} \sum_{i=1}^N p(T_c, \vec{\theta}_i | \vec{D}, I), 
\end{equation}
where $\vec{\theta}_i$ are $N$ random samples drawn from a uniform distribution of EoS parameters in the intervals discussed in Sec.~\ref{sec:eos}, which define the volume $V_{\rm EoS}$. The error can be estimated through the sample variance, and decreases as $1/\sqrt{N}$.

\subsection{Priors}

The joint prior on $T_c$ and $\vec{\theta}$ can be written as
\begin{equation}
p(T_c,\vec{\theta}|I) = p(T_c | \vec{\theta} ,I) p(\vec{\theta}|I), 
\end{equation}
where $p(T_c|\vec{\theta} ,I)$ is the probability that a star with EoS parameters $\vec{\theta}$ has the value $T_c$ for the trace of the energy-momentum tensor at $r=0$, and $p(\vec{\theta}|I)$ is the prior probability on the EoS parameters. This is taken to be uniform in the $\Gamma_i$'s and in $\log (p_1)$ across the intervals discussed in Sec.~\ref{sec:eos}. The parameters are also required to generate a causal EoS that allows a maximum mass larger than $1.95 M_\odot$ (cf.~Sec.~\ref{sec:eos}). The prior $p(T_c | \vec{\theta} ,I)$ on $T_c$ is taken to be uniform in the range of possible values for $T(\rho)$ such that $\rho_{\rm ns} \leq \rho \leq \rho_{\rm max}$. Here $\rho_{\rm ns}$ denotes the nuclear saturation density and $\rho_{\rm \max}$ denotes the central density of the most massive star predicted by the EoS parameters $\vec{\theta}$.

In order to assess the dependence of our results on the EoS prior, we also consider the effect of imposing a maximum mass cutoff.
Recently, several works have attempted to set upper limits to the maximum mass of (nonrotating) neutron stars based on the observation of GW170817 and its electromagnetic counterparts. Combined electromagnetic and gravitational-wave information were used to place the upper limit $M_\textrm{max} \lesssim 2.17 M_\odot$ in Ref.~\cite{Margalit2017}; quasi-universal relations were invoked to derive the constraint $M_\textrm{max} \lesssim 2.16^{+0.17}_{-0.15} M_\odot$ \cite{Rezzolla2018}, and numerical simulations were used to bracket the maximum mass in the $2.15$--$2.28 M_\odot$ interval \cite{Ruiz2018,Shibata2017}. These upper limits also agree with estimates coming from a study of the neutron star mass distribution \cite{Alsing2018}. Therefore, we will additionally explore the effect of a maximum mass cutoff of $2.3 M_\odot$ in our results. 

For each choice of EoS prior, we ensure that that $N$ in Eq.~(\ref{eq:montecarlo}) is sufficiently large that $p(\vec{\theta}|I) \neq 0$ for at least 10,000 samples.

\subsection{Likelihood}

Let us first consider a measurement of the stellar compactness. For simplicity, we assume that the measurement corresponds to a Gaussian distribution around the value predicted by general relativity; it would be straightforward to accommodate any other realistic distribution. Let $C$ be the peak of the distribution determined observationally and $\sigma_C^2$ its variance, which enters as a fixed parameter of the model. With $\vec{D} = \{C\}$, we set
\begin{equation}\label{eq:gaussianC}
p(C | T_c,\vec{\theta},I) = \frac{1}{\sqrt{2\pi \sigma_C^2}} \sum_i e^{-[C - \mathcal{C}_i(T_c,\vec{\theta})]^2/(2\sigma_C^2)}.
\end{equation}
Here $\mathcal{C}_i(T_c,\vec{\theta})$ denotes the stellar compactness obtained by integrating the structure equations with the EoS parameters $\vec{\theta}$ and central density $\rho_{c,i}=\rho_i(T_c)$ [or enthalpy $h_{c,i} = h_i(T_c)$]. The summation index $i$ accounts for the fact that the density may not be a single-valued function of $T$. Indeed, a careful inspection of Fig.~\ref{fig:TcvsC} reveals that a given value of $T_c$ can be associated with more than one stellar compactness.

\begin{figure}[t]
\includegraphics[width=0.48\textwidth]{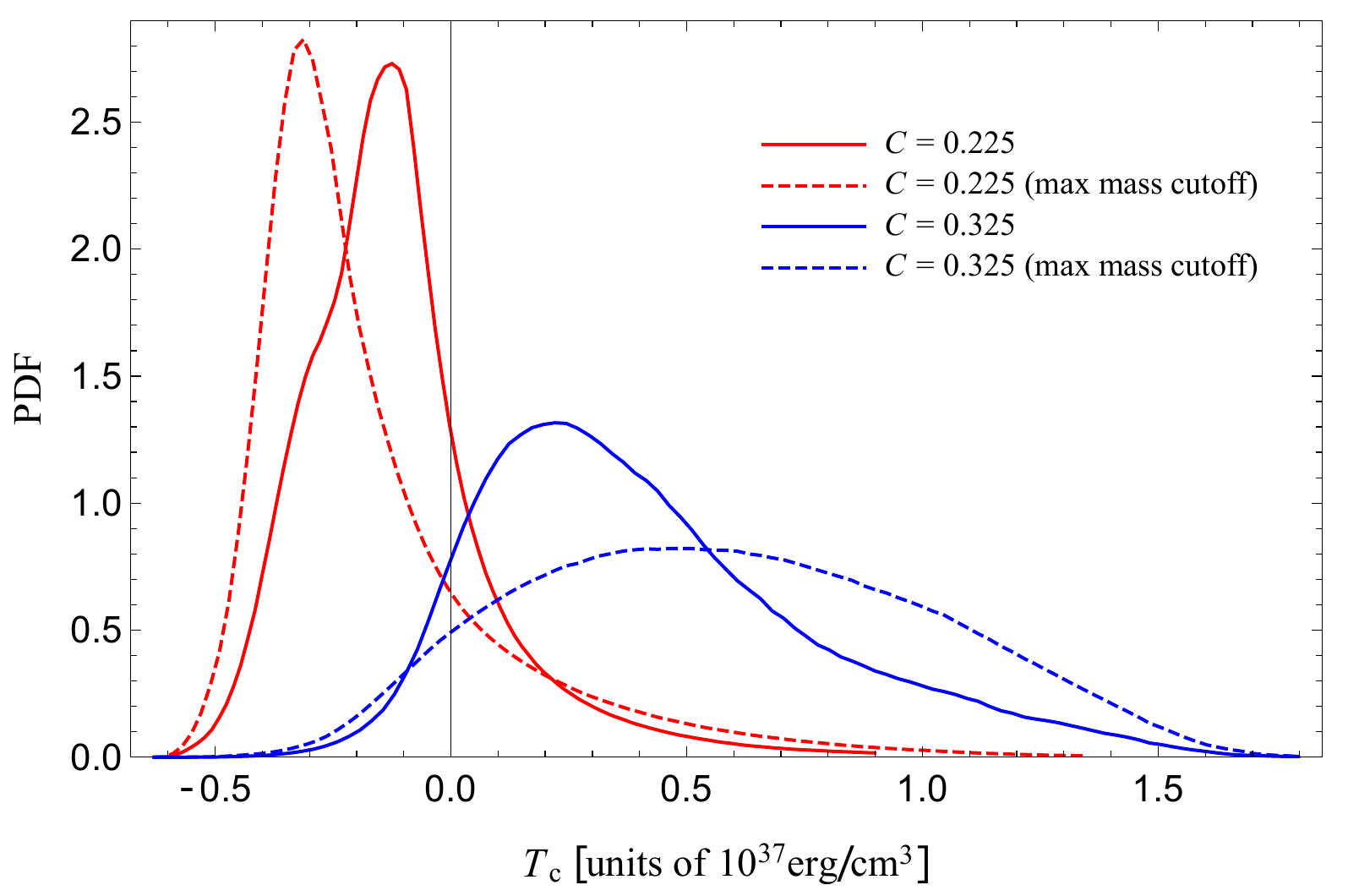}
\caption{Probability distribution functions for $T_c$, for two values of a compactness measurement, $C = 0.225$ and $C=0.325$. For the dashed curves, a maximum-mass cutoff of $2.3M_\odot$ was included in the EoS prior, while no maximum mass cutoff was imposed for the solid curves. In all cases, $\sigma_C = 0.03$. }
\label{fig:massprior}
\end{figure}

We will determine the impact of measuring the mass in addition to the compactness, setting $\vec{D} = \{M,C\}$. If we assume that both the compactness and mass measurements correspond to Gaussian distributions around the theoretical values, we can write
\begin{align}
p(M,C| T_c, \vec{\theta},I) & = \frac{1}{2\pi \sigma_M \sigma_C} \sum_i e^{-[C - \mathcal{C}_i(T_c,\vec{\theta})]^2/(2\sigma_C^2)} \nonumber \\
& \times e^{-[M - \mathcal{M}_i(T_c,\vec{\theta})]^2/(2\sigma_M^2)}.
\end{align}
Here, $\mathcal{M}_i(T_c,\vec{\theta})$ denotes the mass obtained by integrating the structure equations with the EoS parameters $\vec{\theta}$ and central density $\rho_{c,i}=\rho_i(T_c)$, and $\sigma_M$ represents the uncertainty in the mass measurement.

\subsection{Results}

Figure \ref{fig:sigmafixed} shows the probability distribution function (PDF) for the central value of $T$ for several hypothetical measurements of the stellar compactness, ranging from $C = 0.200$ to $C = 0.325$. Here, $\sigma_C$ is fixed to 0.03, which corresponds roughly to a 10\% (or $\sim 1$ km) uncertainty in the radius measurement, assuming a well-measured NS mass. As the compactness increases, it becomes more likely that $T_c>0$, with a 67.6\% probability for $C = 0.275$ and 92.9\% for $C = 0.325$.

Figure \ref{fig:massprior} explores the effect of changing the EoS prior by imposing a maximum mass cutoff of $2.3 M_\odot$. For low compactness, the effect of a maximum mass cutoff is to displace the mean of the distribution towards lower values of $T_c$. For higher values of $C$, the distribution is broadened. For comparison, we get a 67.3\% probability that $T_c>0$ for $C = 0.275$ and 92.6\% for $C = 0.325$; these values are almost the same as those obtained with the less restricted EoS prior.

\begin{figure}[thb]
\includegraphics[width=0.48\textwidth]{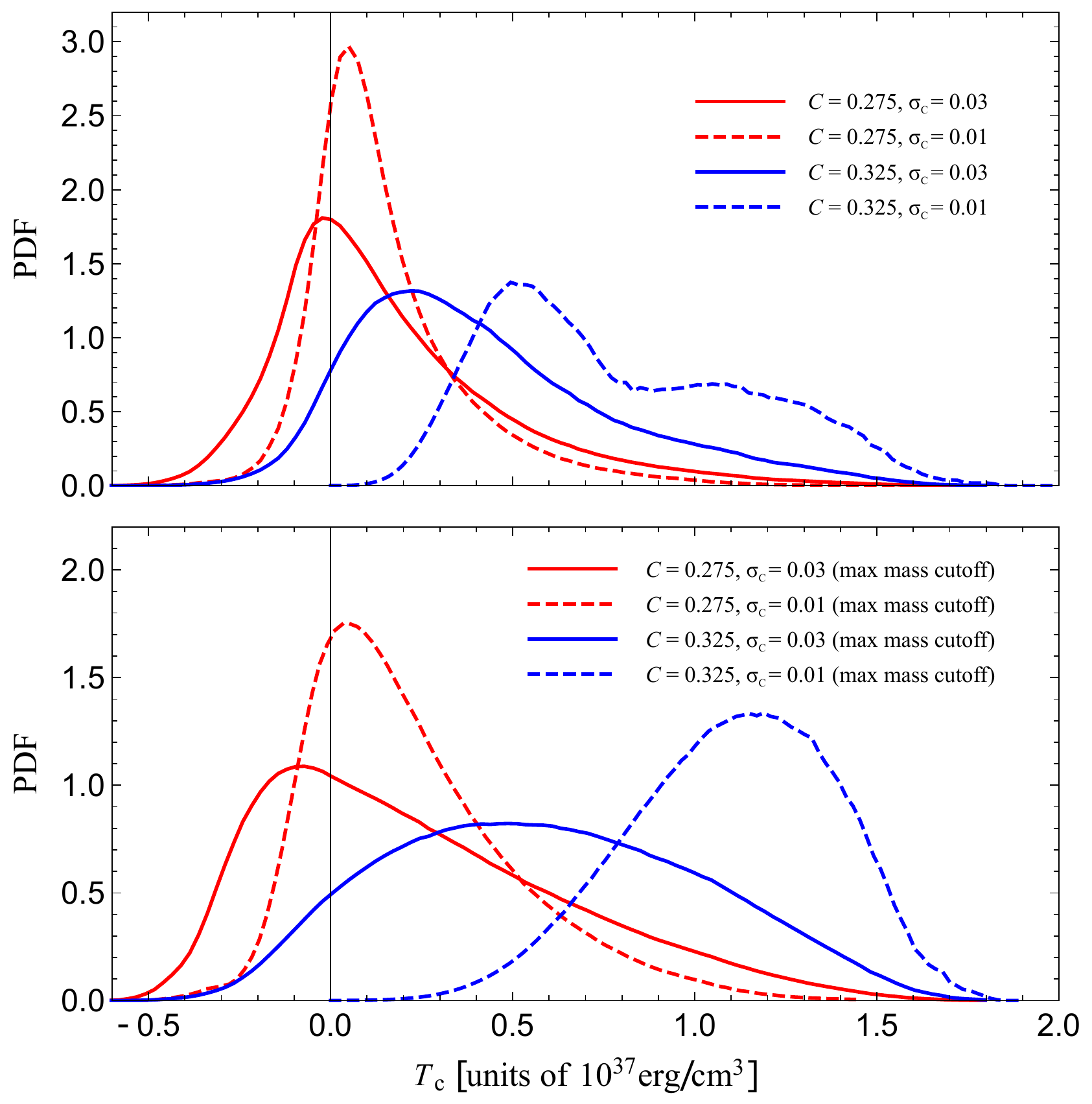}
\caption{Probability distribution functions for $T_c$, for two values of the measured compactness, $C=0.275$ and $C=0.325$, and $\sigma_C = 0.03$ (solid curves) or $\sigma_C = 0.01$ (dashed curves). In the upper panel the PDF is shown for an EoS prior that includes no maximum mass cutoff, while in the lower panel a maximum mass cutoff of $2.3 M_\odot$ is considered.}
\label{fig:Cfixed}
\end{figure}

Figure \ref{fig:Cfixed} shows how the uncertainty $\sigma_C$ in the compactness measurement affects the PDF for $T_c$. Results are displayed for a realistic near-future value $\sigma_C = 0.03$ and a more optimistic value of $\sigma_C = 0.01$. As $\sigma_C$ decreases, the distribution typically becomes narrower and, for higher values of $C$, the PDF moves towards higher values of $T_c$. For comparison, when $\sigma_C = 0.01$ we get a 81.6\% probability that $T_c>0$ for $C = 0.275$, and 100\% for $C=0.325$, using an EoS prior with no maximum-mass cutoff. Incorporating a maximum-mass cutoff alters the shape of the distributions considerably, but the values quoted above remain approximately the same.

\begin{figure}[th]
\includegraphics[width=0.48\textwidth]{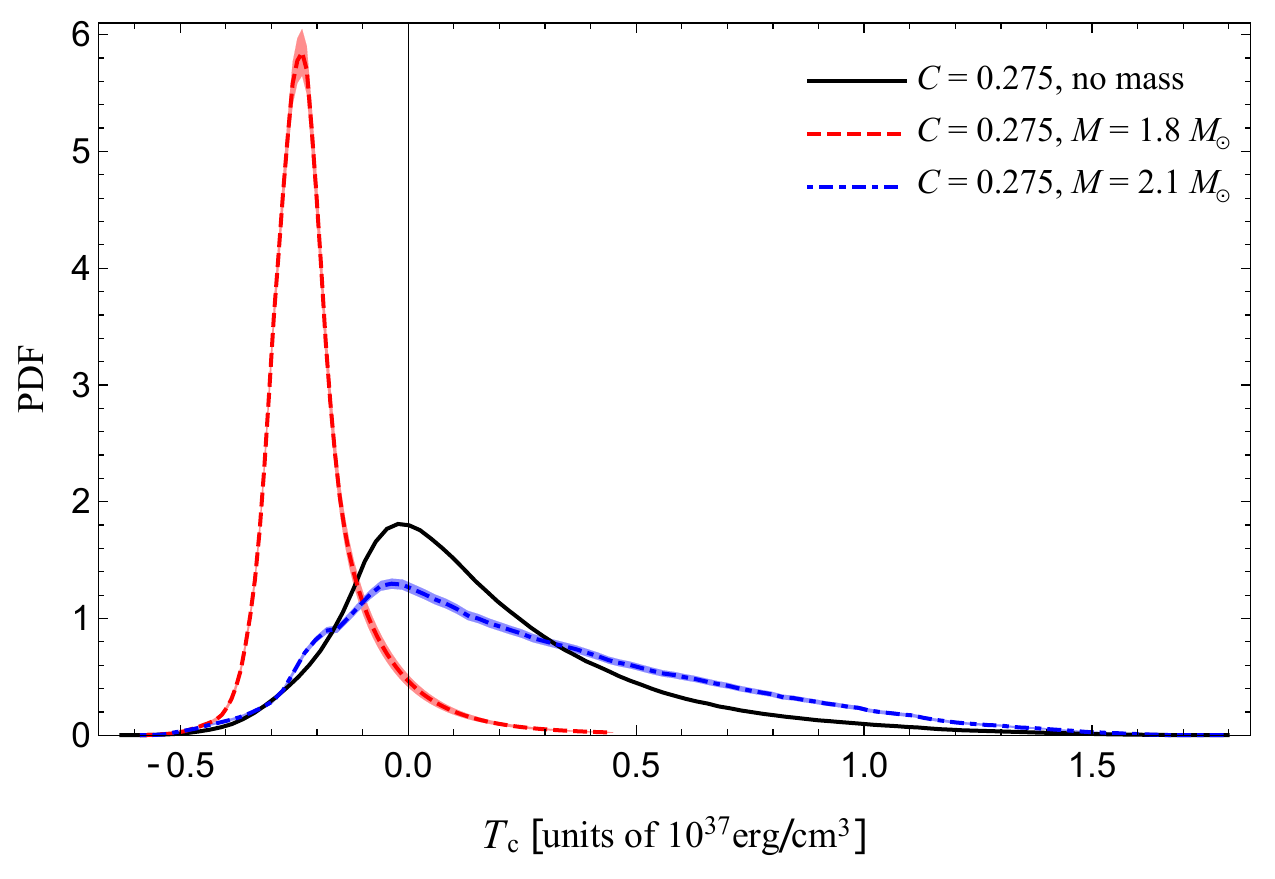}
\caption{Probability distribution function for $T_c$, for a measured stellar compactness of $C=0.275$, with $\sigma_C = 0.03$, and either no concurrent mass measurement or mass measurements of $1.8 M_\odot$ or $2.1 M_\odot$, with $\sigma_M  = 0.05 M_\odot$. Shades represent the error in the marginalization procedure.}
\label{fig:combinedwithmass}
\end{figure}

In Fig.~\ref{fig:combinedwithmass} we show how a simultaneous measurement of the mass changes the probability distribution for $T_c$. We consider $C=0.275$, $\sigma_C = 0.03$, and fix the uncertainty in the mass measurement to be small, given by $\sigma_M = 0.05 M_\odot$. For such a high compactness, a large value for the measured mass (such as $2.1 M_\odot$ in this example) is essentially uninformative, and the distribution is roughly unchanged. However, a relatively small mass (such as $1.8 M_\odot$ in this example) dramatically moves the distribution towards negative values of $T_c$. This illustrates the fact, already visible in Fig.~\ref{fig:massradius}, that although a large mass is not a requirement for a neutron star to feature a positive $T$, the condition $T_c>0$ only occurs for the most massive stars predicted by each EoS.

\section{Conclusions} \label{sec:conclusions}

Scalar-tensor theories of gravity predict a rich phenomenology for neutrons stars when $T = 3p-\epsilon$ becomes positive in a region of the stellar interior. In this work we investigated the relation between this microscopic feature---which depends on the yet-unknown behavior of the equation of state at supranuclear densities---and macroscopic, observable properties of neutron stars. 

We found that the configuration along a sequence of equilibrium solutions at which $T$ first becomes positive (at the stellar center) has a quasi-universal compactness, given by $C_{T_c = 0} = 0.262^{+0.011}_{-0.017}$ (cf.~Figs.~\ref{fig:massradius} and \ref{fig:histogramC}). For a star in the mass range $1.97$--$2.05 M_\odot$, this translates into a radius of $11.2^{+1.0}_{-0.5}$ km, which is consistent with radii estimates built from spectroscopic measurements \cite{Ozel2016}. 

The positiveness of the trace of the energy-momentum tensor is also related to the condition $c_s > c/\sqrt{3}$, which has been discussed recently in the literature \cite{Bedaque2015,Alsing2018}. Indeed, for a linear equation of state of the form $p = \alpha \epsilon$, these conditions are equivalent, but they differ for nonlinear EoS (cf.~Fig.~\ref{fig:histogramcs}). In Refs.~\cite{Bedaque2015,Alsing2018} it was shown that the observed high masses of some neutron stars rule out $c_s < c/\sqrt{3}$---a condition that had been advocated on theoretical grounds---with high significance. Similarly, the observed high masses and relatively low radii of neutron stars make it plausible that the condition $T>0$ is realized inside the most compact and massive neutron stars in Nature.

Measurements of neutron-star radii are still imprecise, and so are direct estimates of the stellar compactness (through, for example, the redshift of atomic spectral lines). In this paper we have imagined a proximate future in which a number of compactness measurements have been obtained with uncertainties no larger than approximately $10\%$, and explored the impact of such measurements on the determination of $T_c$, the central value of the trace of the energy-momentum tensor for the nuclear matter that makes up a neutron star. Our main results are summarized in Figs.~\ref{fig:sigmafixed}, \ref{fig:massprior}, \ref{fig:Cfixed}, and \ref{fig:combinedwithmass}. An observation of a neutron star with a high compactness and a large mass will allow us to infer with high confidence that $T_c > 0$, in spite of our rudimentary knowledge of the equation of state. This star will then constitute a unique laboratory to probe a wide class of scalar extensions to general relativity \cite{Mendes2015,Mendes2016}, which would predict properties that are dramatically different from those of a general-relativistic star.

\acknowledgments
R.~M.~is grateful to R.~Negreiros for insightful discussions on the nuclear EoS. 
The work carried out at the University of Guelph was supported by the Natural Sciences and Engineering Research Council of Canada. 

\bibliography{references}

\begin{thebibliography}{37}%
\makeatletter
\providecommand \@ifxundefined [1]{%
 \@ifx{#1\undefined}
}%
\providecommand \@ifnum [1]{%
 \ifnum #1\expandafter \@firstoftwo
 \else \expandafter \@secondoftwo
 \fi
}%
\providecommand \@ifx [1]{%
 \ifx #1\expandafter \@firstoftwo
 \else \expandafter \@secondoftwo
 \fi
}%
\providecommand \natexlab [1]{#1}%
\providecommand \enquote  [1]{``#1''}%
\providecommand \bibnamefont  [1]{#1}%
\providecommand \bibfnamefont [1]{#1}%
\providecommand \citenamefont [1]{#1}%
\providecommand \href@noop [0]{\@secondoftwo}%
\providecommand \href [0]{\begingroup \@sanitize@url \@href}%
\providecommand \@href[1]{\@@startlink{#1}\@@href}%
\providecommand \@@href[1]{\endgroup#1\@@endlink}%
\providecommand \@sanitize@url [0]{\catcode `\\12\catcode `\$12\catcode
  `\&12\catcode `\#12\catcode `\^12\catcode `\_12\catcode `\%12\relax}%
\providecommand \@@startlink[1]{}%
\providecommand \@@endlink[0]{}%
\providecommand \url  [0]{\begingroup\@sanitize@url \@url }%
\providecommand \@url [1]{\endgroup\@href {#1}{\urlprefix }}%
\providecommand \urlprefix  [0]{URL }%
\providecommand \Eprint [0]{\href }%
\providecommand \doibase [0]{http://dx.doi.org/}%
\providecommand \selectlanguage [0]{\@gobble}%
\providecommand \bibinfo  [0]{\@secondoftwo}%
\providecommand \bibfield  [0]{\@secondoftwo}%
\providecommand \translation [1]{[#1]}%
\providecommand \BibitemOpen [0]{}%
\providecommand \bibitemStop [0]{}%
\providecommand \bibitemNoStop [0]{.\EOS\space}%
\providecommand \EOS [0]{\spacefactor3000\relax}%
\providecommand \BibitemShut  [1]{\csname bibitem#1\endcsname}%
\let\auto@bib@innerbib\@empty
\bibitem [{\citenamefont {Damour}\ and\ \citenamefont
  {Esposito-Far\`ese}(1992)}]{Damour1992}%
  \BibitemOpen
  \bibfield  {author} {\bibinfo {author} {\bibfnamefont {T.}~\bibnamefont
  {Damour}}\ and\ \bibinfo {author} {\bibfnamefont {G.}~\bibnamefont
  {Esposito-Far\`ese}},\ }\href {\doibase 10.1088/0264-9381/9/9/015} {\bibfield
   {journal} {\bibinfo  {journal} {Classical Quantum Gravity}\ }\textbf
  {\bibinfo {volume} {9}},\ \bibinfo {pages} {2093} (\bibinfo {year}
  {1992})}\BibitemShut {NoStop}%
\bibitem [{\citenamefont {Mendes}(2015)}]{Mendes2015}%
  \BibitemOpen
  \bibfield  {author} {\bibinfo {author} {\bibfnamefont {R.~F.~P.}\
  \bibnamefont {Mendes}},\ }\href {\doibase 10.1103/PhysRevD.91.064024}
  {\bibfield  {journal} {\bibinfo  {journal} {Phys. Rev. D}\ }\textbf {\bibinfo
  {volume} {91}},\ \bibinfo {pages} {064024} (\bibinfo {year}
  {2015})}\BibitemShut {NoStop}%
\bibitem [{\citenamefont {Palenzuela}\ and\ \citenamefont
  {Liebling}(2016)}]{Palenzuela2016}%
  \BibitemOpen
  \bibfield  {author} {\bibinfo {author} {\bibfnamefont {C.}~\bibnamefont
  {Palenzuela}}\ and\ \bibinfo {author} {\bibfnamefont {S.}~\bibnamefont
  {Liebling}},\ }\href@noop {} {\bibfield  {journal} {\bibinfo  {journal}
  {Phys. Rev. D}\ }\textbf {\bibinfo {volume} {93}},\ \bibinfo {pages} {044009}
  (\bibinfo {year} {2016})}\BibitemShut {NoStop}%
\bibitem [{\citenamefont {Mendes}\ and\ \citenamefont
  {Ortiz}(2016)}]{Mendes2016}%
  \BibitemOpen
  \bibfield  {author} {\bibinfo {author} {\bibfnamefont {R.~F.~P.}\
  \bibnamefont {Mendes}}\ and\ \bibinfo {author} {\bibfnamefont
  {N.}~\bibnamefont {Ortiz}},\ }\href {\doibase 10.1103/PhysRevD.93.124035}
  {\bibfield  {journal} {\bibinfo  {journal} {Phys. Rev. D}\ }\textbf {\bibinfo
  {volume} {93}},\ \bibinfo {pages} {124035} (\bibinfo {year}
  {2016})}\BibitemShut {NoStop}%
\bibitem [{\citenamefont {Damour}\ and\ \citenamefont
  {Esposito-Far{\`{e}}se}(1996)}]{Damour1996a}%
  \BibitemOpen
  \bibfield  {author} {\bibinfo {author} {\bibfnamefont {T.}~\bibnamefont
  {Damour}}\ and\ \bibinfo {author} {\bibfnamefont {G.}~\bibnamefont
  {Esposito-Far{\`{e}}se}},\ }\href {\doibase 10.1103/PhysRevD.53.5541}
  {\bibfield  {journal} {\bibinfo  {journal} {Phys. Rev. D}\ }\textbf {\bibinfo
  {volume} {53}},\ \bibinfo {pages} {5541} (\bibinfo {year}
  {1996})}\BibitemShut {NoStop}%
\bibitem [{\citenamefont {Anderson}\ and\ \citenamefont
  {Yunes}(2017)}]{Anderson2017}%
  \BibitemOpen
  \bibfield  {author} {\bibinfo {author} {\bibfnamefont {D.}~\bibnamefont
  {Anderson}}\ and\ \bibinfo {author} {\bibfnamefont {N.}~\bibnamefont
  {Yunes}},\ }\href {\doibase 10.1103/PhysRevD.96.064037} {\bibfield  {journal}
  {\bibinfo  {journal} {Phys. Rev. D}\ }\textbf {\bibinfo {volume} {96}},\
  \bibinfo {pages} {064037} (\bibinfo {year} {2017})}\BibitemShut {NoStop}%
\bibitem [{\citenamefont {Damour}\ and\ \citenamefont
  {Esposito-Far{\`{e}}se}(1993)}]{Damour1993}%
  \BibitemOpen
  \bibfield  {author} {\bibinfo {author} {\bibfnamefont {T.}~\bibnamefont
  {Damour}}\ and\ \bibinfo {author} {\bibfnamefont {G.}~\bibnamefont
  {Esposito-Far{\`{e}}se}},\ }\href {\doibase 10.1103/PhysRevLett.70.2220}
  {\bibfield  {journal} {\bibinfo  {journal} {Phys. Rev. Lett.}\ }\textbf
  {\bibinfo {volume} {70}},\ \bibinfo {pages} {2220} (\bibinfo {year}
  {1993})}\BibitemShut {NoStop}%
\bibitem [{\citenamefont {Babichev}\ and\ \citenamefont
  {Langlois}(2010)}]{Babichev2010}%
  \BibitemOpen
  \bibfield  {author} {\bibinfo {author} {\bibfnamefont {E.}~\bibnamefont
  {Babichev}}\ and\ \bibinfo {author} {\bibfnamefont {D.}~\bibnamefont
  {Langlois}},\ }\href {\doibase 10.1103/PhysRevD.81.124051} {\bibfield
  {journal} {\bibinfo  {journal} {Phys. Rev. D}\ }\textbf {\bibinfo {volume}
  {81}},\ \bibinfo {pages} {124051} (\bibinfo {year} {2010})}\BibitemShut
  {NoStop}%
\bibitem [{\citenamefont {Brax}\ \emph {et~al.}(2017)\citenamefont {Brax},
  \citenamefont {Davis},\ and\ \citenamefont {Jha}}]{Brax2017}%
  \BibitemOpen
  \bibfield  {author} {\bibinfo {author} {\bibfnamefont {P.}~\bibnamefont
  {Brax}}, \bibinfo {author} {\bibfnamefont {A.-C.}\ \bibnamefont {Davis}}, \
  and\ \bibinfo {author} {\bibfnamefont {R.}~\bibnamefont {Jha}},\ }\href
  {\doibase 10.1103/PhysRevD.95.083514} {\bibfield  {journal} {\bibinfo
  {journal} {Phys. Rev. D}\ }\textbf {\bibinfo {volume} {95}},\ \bibinfo
  {pages} {083514} (\bibinfo {year} {2017})}\BibitemShut {NoStop}%
\bibitem [{\citenamefont {Landau}\ and\ \citenamefont
  {Lifshitz}(1975)}]{Landau1951}%
  \BibitemOpen
  \bibfield  {author} {\bibinfo {author} {\bibfnamefont {L.~D.}\ \bibnamefont
  {Landau}}\ and\ \bibinfo {author} {\bibfnamefont {E.~M.}\ \bibnamefont
  {Lifshitz}},\ }\href@noop {} {\emph {\bibinfo {title} {The Classical Theory
  of Fields}}}\ (\bibinfo  {publisher} {Reed Educational and Professional
  Publishing Ltd},\ \bibinfo {address} {Oxford, UK},\ \bibinfo {year}
  {1975})\BibitemShut {NoStop}%
\bibitem [{\citenamefont {Zeldovich}(1962)}]{Zeldovich1962}%
  \BibitemOpen
  \bibfield  {author} {\bibinfo {author} {\bibfnamefont {Y.~B.}\ \bibnamefont
  {Zeldovich}},\ }\href@noop {} {\bibfield  {journal} {\bibinfo  {journal} {J.
  Exp. Theor. Phys.}\ }\textbf {\bibinfo {volume} {14}},\ \bibinfo {pages}
  {1143} (\bibinfo {year} {1962})}\BibitemShut {NoStop}%
\bibitem [{\citenamefont {Haensel}\ \emph {et~al.}(2007)\citenamefont
  {Haensel}, \citenamefont {Potekhin},\ and\ \citenamefont
  {Yakovlev}}]{Haensel2007}%
  \BibitemOpen
  \bibfield  {author} {\bibinfo {author} {\bibfnamefont {P.}~\bibnamefont
  {Haensel}}, \bibinfo {author} {\bibfnamefont {A.~Y.}\ \bibnamefont
  {Potekhin}}, \ and\ \bibinfo {author} {\bibfnamefont {D.~G.}\ \bibnamefont
  {Yakovlev}},\ }\href@noop {} {\emph {\bibinfo {title} {Neutron stars 1:
  Equation of state and structure}}}\ (\bibinfo  {publisher} {Springer},\
  \bibinfo {year} {2007})\BibitemShut {NoStop}%
\bibitem [{\citenamefont {\"{O}zel}\ and\ \citenamefont
  {Freire}(2016)}]{Ozel2016}%
  \BibitemOpen
  \bibfield  {author} {\bibinfo {author} {\bibfnamefont {F.}~\bibnamefont
  {\"{O}zel}}\ and\ \bibinfo {author} {\bibfnamefont {P.}~\bibnamefont
  {Freire}},\ }\href {\doibase 10.1146/annurev-astro-081915-023322} {\bibfield
  {journal} {\bibinfo  {journal} {Annual Review of Astronomy and Astrophysics}\
  }\textbf {\bibinfo {volume} {54}},\ \bibinfo {pages} {401} (\bibinfo {year}
  {2016})}\BibitemShut {NoStop}%
\bibitem [{\citenamefont {Cottam}\ \emph {et~al.}(2002)\citenamefont {Cottam},
  \citenamefont {Paerels},\ and\ \citenamefont {M\'{e}ndez}}]{Cottam2002}%
  \BibitemOpen
  \bibfield  {author} {\bibinfo {author} {\bibfnamefont {J.}~\bibnamefont
  {Cottam}}, \bibinfo {author} {\bibfnamefont {F.}~\bibnamefont {Paerels}}, \
  and\ \bibinfo {author} {\bibfnamefont {M.}~\bibnamefont {M\'{e}ndez}},\
  }\href {\doibase 10.1038/nature01159} {\bibfield  {journal} {\bibinfo
  {journal} {Nature}\ }\textbf {\bibinfo {volume} {420}},\ \bibinfo {pages}
  {51} (\bibinfo {year} {2002})}\BibitemShut {NoStop}%
\bibitem [{\citenamefont {Cottam}\ \emph {et~al.}(2008)\citenamefont {Cottam},
  \citenamefont {Paerels}, \citenamefont {M\'{e}ndez}, \citenamefont {Boirin},
  \citenamefont {Lewin}, \citenamefont {Kuulkers},\ and\ \citenamefont
  {Miller}}]{Cottam2008}%
  \BibitemOpen
  \bibfield  {author} {\bibinfo {author} {\bibfnamefont {J.}~\bibnamefont
  {Cottam}}, \bibinfo {author} {\bibfnamefont {F.}~\bibnamefont {Paerels}},
  \bibinfo {author} {\bibfnamefont {M.}~\bibnamefont {M\'{e}ndez}}, \bibinfo
  {author} {\bibfnamefont {L.}~\bibnamefont {Boirin}}, \bibinfo {author}
  {\bibfnamefont {W.~H.~G.}\ \bibnamefont {Lewin}}, \bibinfo {author}
  {\bibfnamefont {E.}~\bibnamefont {Kuulkers}}, \ and\ \bibinfo {author}
  {\bibfnamefont {J.~M.}\ \bibnamefont {Miller}},\ }\href
  {http://stacks.iop.org/0004-637X/672/i=1/a=504} {\bibfield  {journal}
  {\bibinfo  {journal} {The Astrophysical Journal}\ }\textbf {\bibinfo {volume}
  {672}},\ \bibinfo {pages} {504} (\bibinfo {year} {2008})}\BibitemShut
  {NoStop}%
\bibitem [{\citenamefont {Gendreau}\ \emph {et~al.}(2016)\citenamefont
  {Gendreau} \emph {et~al.}}]{Gendreau2016}%
  \BibitemOpen
  \bibfield  {author} {\bibinfo {author} {\bibfnamefont {K.~C.}\ \bibnamefont
  {Gendreau}} \emph {et~al.},\ }\href {\doibase 10.1117/12.2231304} {\bibfield
  {journal} {\bibinfo  {journal} {Proc. of SPIE}\ }\textbf {\bibinfo {volume}
  {9905}},\ \bibinfo {pages} {99051H} (\bibinfo {year} {2016})}\BibitemShut
  {NoStop}%
\bibitem [{\citenamefont {{Feroci}}\ \emph {et~al.}(2012)\citenamefont
  {{Feroci}} \emph {et~al.}}]{Feroci2012}%
  \BibitemOpen
  \bibfield  {author} {\bibinfo {author} {\bibfnamefont {M.}~\bibnamefont
  {{Feroci}}} \emph {et~al.},\ }\href {\doibase 10.1007/s10686-011-9237-2}
  {\bibfield  {journal} {\bibinfo  {journal} {Experimental Astronomy}\ }\textbf
  {\bibinfo {volume} {34}},\ \bibinfo {pages} {415} (\bibinfo {year}
  {2012})}\BibitemShut {NoStop}%
\bibitem [{\citenamefont {Raithel}\ \emph
  {et~al.}(2016{\natexlab{a}})\citenamefont {Raithel}, \citenamefont
  {{\"{O}}zel},\ and\ \citenamefont {Psaltis}}]{Raithel2016}%
  \BibitemOpen
  \bibfield  {author} {\bibinfo {author} {\bibfnamefont {C.~A.}\ \bibnamefont
  {Raithel}}, \bibinfo {author} {\bibfnamefont {F.}~\bibnamefont {{\"{O}}zel}},
  \ and\ \bibinfo {author} {\bibfnamefont {D.}~\bibnamefont {Psaltis}},\ }\href
  {\doibase 10.1103/PhysRevC.93.032801} {\bibfield  {journal} {\bibinfo
  {journal} {Phys. Rev. C}\ }\textbf {\bibinfo {volume} {93}},\ \bibinfo
  {pages} {032801} (\bibinfo {year} {2016}{\natexlab{a}})}\BibitemShut
  {NoStop}%
\bibitem [{\citenamefont {{LIGO Scientific Collaboration}}\ and\ \citenamefont
  {{VIRGO Collaboration}}(2017)}]{Abbott2017a}%
  \BibitemOpen
  \bibfield  {author} {\bibinfo {author} {\bibnamefont {{LIGO Scientific
  Collaboration}}}\ and\ \bibinfo {author} {\bibnamefont {{VIRGO
  Collaboration}}},\ }\href {\doibase 10.1103/PhysRevLett.119.161101}
  {\bibfield  {journal} {\bibinfo  {journal} {Phys. Rev. Lett.}\ }\textbf
  {\bibinfo {volume} {119}},\ \bibinfo {pages} {161101} (\bibinfo {year}
  {2017})}\BibitemShut {NoStop}%
\bibitem [{\citenamefont {{LIGO Scientific Collaboration}}\ and\ \citenamefont
  {{Virgo Collaboration}}(2018)}]{LIGO2018}%
  \BibitemOpen
  \bibfield  {author} {\bibinfo {author} {\bibnamefont {{LIGO Scientific
  Collaboration}}}\ and\ \bibinfo {author} {\bibnamefont {{Virgo
  Collaboration}}},\ }\href@noop {} {\bibfield  {journal} {\bibinfo  {journal}
  {ArXiv e-prints}\ } (\bibinfo {year} {2018})},\ \Eprint
  {http://arxiv.org/abs/1805.11581} {arXiv:1805.11581 [gr-qc]} \BibitemShut
  {NoStop}%
\bibitem [{\citenamefont {{\"{O}}zel}\ \emph {et~al.}(2016)\citenamefont
  {{\"{O}}zel}, \citenamefont {Psaltis}, \citenamefont {G\"{u}ver},
  \citenamefont {Baym}, \citenamefont {Heinke},\ and\ \citenamefont
  {Guillot}}]{Ozel2016a}%
  \BibitemOpen
  \bibfield  {author} {\bibinfo {author} {\bibfnamefont {F.}~\bibnamefont
  {{\"{O}}zel}}, \bibinfo {author} {\bibfnamefont {D.}~\bibnamefont {Psaltis}},
  \bibinfo {author} {\bibfnamefont {T.}~\bibnamefont {G\"{u}ver}}, \bibinfo
  {author} {\bibfnamefont {G.}~\bibnamefont {Baym}}, \bibinfo {author}
  {\bibfnamefont {C.}~\bibnamefont {Heinke}}, \ and\ \bibinfo {author}
  {\bibfnamefont {S.}~\bibnamefont {Guillot}},\ }\href
  {http://stacks.iop.org/0004-637X/820/i=1/a=28} {\bibfield  {journal}
  {\bibinfo  {journal} {The Astrophysical Journal}\ }\textbf {\bibinfo {volume}
  {820}},\ \bibinfo {pages} {28} (\bibinfo {year} {2016})}\BibitemShut
  {NoStop}%
\bibitem [{\citenamefont {Read}\ \emph {et~al.}(2009)\citenamefont {Read},
  \citenamefont {Lackey}, \citenamefont {Owen},\ and\ \citenamefont
  {Friedman}}]{Read2009}%
  \BibitemOpen
  \bibfield  {author} {\bibinfo {author} {\bibfnamefont {J.}~\bibnamefont
  {Read}}, \bibinfo {author} {\bibfnamefont {B.}~\bibnamefont {Lackey}},
  \bibinfo {author} {\bibfnamefont {B.}~\bibnamefont {Owen}}, \ and\ \bibinfo
  {author} {\bibfnamefont {J.~L.}\ \bibnamefont {Friedman}},\ }\href {\doibase
  10.1103/PhysRevD.79.124032} {\bibfield  {journal} {\bibinfo  {journal} {Phys.
  Rev. D}\ }\textbf {\bibinfo {volume} {79}},\ \bibinfo {pages} {124032}
  (\bibinfo {year} {2009})}\BibitemShut {NoStop}%
\bibitem [{\citenamefont {Steiner}\ \emph {et~al.}(2010)\citenamefont
  {Steiner}, \citenamefont {Lattimer},\ and\ \citenamefont
  {Brown}}]{Steiner2010}%
  \BibitemOpen
  \bibfield  {author} {\bibinfo {author} {\bibfnamefont {A.~W.}\ \bibnamefont
  {Steiner}}, \bibinfo {author} {\bibfnamefont {J.~M.}\ \bibnamefont
  {Lattimer}}, \ and\ \bibinfo {author} {\bibfnamefont {E.~F.}\ \bibnamefont
  {Brown}},\ }\href {\doibase 10.1088/0004-637X/722/1/33} {\bibfield  {journal}
  {\bibinfo  {journal} {Astrophys. J.}\ }\textbf {\bibinfo {volume} {722}},\
  \bibinfo {pages} {33} (\bibinfo {year} {2010})}\BibitemShut {NoStop}%
\bibitem [{\citenamefont {Hebeler}\ \emph {et~al.}(2013)\citenamefont
  {Hebeler}, \citenamefont {Lattimer}, \citenamefont {Pethick},\ and\
  \citenamefont {Schwenk}}]{Hebeler2013}%
  \BibitemOpen
  \bibfield  {author} {\bibinfo {author} {\bibfnamefont {K.}~\bibnamefont
  {Hebeler}}, \bibinfo {author} {\bibfnamefont {J.~M.}\ \bibnamefont
  {Lattimer}}, \bibinfo {author} {\bibfnamefont {C.~J.}\ \bibnamefont
  {Pethick}}, \ and\ \bibinfo {author} {\bibfnamefont {A.}~\bibnamefont
  {Schwenk}},\ }\href {\doibase 10.1088/0004-637X/773/1/11} {\bibfield
  {journal} {\bibinfo  {journal} {Astrophys. J.}\ }\textbf {\bibinfo {volume}
  {773}},\ \bibinfo {pages} {11} (\bibinfo {year} {2013})}\BibitemShut
  {NoStop}%
\bibitem [{\citenamefont {Steiner}\ \emph {et~al.}(2013)\citenamefont
  {Steiner}, \citenamefont {Lattimer},\ and\ \citenamefont
  {Brown}}]{Steiner2013}%
  \BibitemOpen
  \bibfield  {author} {\bibinfo {author} {\bibfnamefont {A.~W.}\ \bibnamefont
  {Steiner}}, \bibinfo {author} {\bibfnamefont {J.~M.}\ \bibnamefont
  {Lattimer}}, \ and\ \bibinfo {author} {\bibfnamefont {E.~F.}\ \bibnamefont
  {Brown}},\ }\href {\doibase 10.1088/2041-8205/765/1/L5} {\bibfield  {journal}
  {\bibinfo  {journal} {Astrophys. J.}\ }\textbf {\bibinfo {volume} {765}},\
  \bibinfo {pages} {L5} (\bibinfo {year} {2013})}\BibitemShut {NoStop}%
\bibitem [{\citenamefont {Raithel}\ \emph
  {et~al.}(2016{\natexlab{b}})\citenamefont {Raithel}, \citenamefont
  {{\"{O}}zel},\ and\ \citenamefont {Psaltis}}]{Raithel2016a}%
  \BibitemOpen
  \bibfield  {author} {\bibinfo {author} {\bibfnamefont {C.~A.}\ \bibnamefont
  {Raithel}}, \bibinfo {author} {\bibfnamefont {F.}~\bibnamefont {{\"{O}}zel}},
  \ and\ \bibinfo {author} {\bibfnamefont {D.}~\bibnamefont {Psaltis}},\ }\href
  {\doibase 10.3847/0004-637X/831/1/44} {\bibfield  {journal} {\bibinfo
  {journal} {Astrophys. J.}\ }\textbf {\bibinfo {volume} {831}},\ \bibinfo
  {pages} {44} (\bibinfo {year} {2016}{\natexlab{b}})}\BibitemShut {NoStop}%
\bibitem [{\citenamefont {Raaijmakers}\ \emph {et~al.}(2018)\citenamefont
  {Raaijmakers}, \citenamefont {Riley},\ and\ \citenamefont
  {Watts}}]{Raaijmakers2018}%
  \BibitemOpen
  \bibfield  {author} {\bibinfo {author} {\bibfnamefont {G.}~\bibnamefont
  {Raaijmakers}}, \bibinfo {author} {\bibfnamefont {T.~E.}\ \bibnamefont
  {Riley}}, \ and\ \bibinfo {author} {\bibfnamefont {A.~L.}\ \bibnamefont
  {Watts}},\ }\href {\doibase 10.1093/mnras/sty1052} {\bibfield  {journal}
  {\bibinfo  {journal} {Monthly Notices of the Royal Astronomical Society}\
  }\textbf {\bibinfo {volume} {478}},\ \bibinfo {pages} {2177} (\bibinfo {year}
  {2018})}\BibitemShut {NoStop}%
\bibitem [{\citenamefont {Douchin}\ and\ \citenamefont
  {Haensel}(2001)}]{Douchin2001}%
  \BibitemOpen
  \bibfield  {author} {\bibinfo {author} {\bibfnamefont {F.}~\bibnamefont
  {Douchin}}\ and\ \bibinfo {author} {\bibfnamefont {P.}~\bibnamefont
  {Haensel}},\ }\href@noop {} {\bibfield  {journal} {\bibinfo  {journal}
  {Astron. Astrophys}\ }\textbf {\bibinfo {volume} {380}},\ \bibinfo {pages}
  {151} (\bibinfo {year} {2001})}\BibitemShut {NoStop}%
\bibitem [{\citenamefont {Antoniadis}\ \emph {et~al.}(2013)\citenamefont
  {Antoniadis} \emph {et~al.}}]{Antoniadis2013}%
  \BibitemOpen
  \bibfield  {author} {\bibinfo {author} {\bibfnamefont {J.}~\bibnamefont
  {Antoniadis}} \emph {et~al.},\ }\href {\doibase 10.1126/science.1233232}
  {\bibfield  {journal} {\bibinfo  {journal} {Science}\ }\textbf {\bibinfo
  {volume} {340}},\ \bibinfo {pages} {1233232} (\bibinfo {year}
  {2013})}\BibitemShut {NoStop}%
\bibitem [{\citenamefont {Lindblom}(1992)}]{Lindblom1992a}%
  \BibitemOpen
  \bibfield  {author} {\bibinfo {author} {\bibfnamefont {L.}~\bibnamefont
  {Lindblom}},\ }\href {\doibase 10.1086/171882} {\bibfield  {journal}
  {\bibinfo  {journal} {Astrophys. J.}\ }\textbf {\bibinfo {volume} {398}},\
  \bibinfo {pages} {569} (\bibinfo {year} {1992})}\BibitemShut {NoStop}%
\bibitem [{\citenamefont {Guillot}\ and\ \citenamefont
  {Rutledge}(2014)}]{Guillot2014}%
  \BibitemOpen
  \bibfield  {author} {\bibinfo {author} {\bibfnamefont {S.}~\bibnamefont
  {Guillot}}\ and\ \bibinfo {author} {\bibfnamefont {R.~E.}\ \bibnamefont
  {Rutledge}},\ }\href {http://stacks.iop.org/2041-8205/796/i=1/a=L3}
  {\bibfield  {journal} {\bibinfo  {journal} {The Astrophysical Journal
  Letters}\ }\textbf {\bibinfo {volume} {796}},\ \bibinfo {pages} {L3}
  (\bibinfo {year} {2014})}\BibitemShut {NoStop}%
\bibitem [{\citenamefont {Margalit}\ and\ \citenamefont
  {Metzger}(2017)}]{Margalit2017}%
  \BibitemOpen
  \bibfield  {author} {\bibinfo {author} {\bibfnamefont {B.}~\bibnamefont
  {Margalit}}\ and\ \bibinfo {author} {\bibfnamefont {B.~D.}\ \bibnamefont
  {Metzger}},\ }\href {\doibase 10.3847/2041-8213/aa991c} {\bibfield  {journal}
  {\bibinfo  {journal} {Astrophys. J.}\ }\textbf {\bibinfo {volume} {850}},\
  \bibinfo {pages} {L19} (\bibinfo {year} {2017})}\BibitemShut {NoStop}%
\bibitem [{\citenamefont {Rezzolla}\ \emph {et~al.}(2018)\citenamefont
  {Rezzolla}, \citenamefont {Most},\ and\ \citenamefont {Weih}}]{Rezzolla2018}%
  \BibitemOpen
  \bibfield  {author} {\bibinfo {author} {\bibfnamefont {L.}~\bibnamefont
  {Rezzolla}}, \bibinfo {author} {\bibfnamefont {E.~R.}\ \bibnamefont {Most}},
  \ and\ \bibinfo {author} {\bibfnamefont {L.~R.}\ \bibnamefont {Weih}},\
  }\href {\doibase 10.3847/2041-8213/aaa401} {\bibfield  {journal} {\bibinfo
  {journal} {Astrophys. J.}\ }\textbf {\bibinfo {volume} {852}},\ \bibinfo
  {pages} {L25} (\bibinfo {year} {2018})}\BibitemShut {NoStop}%
\bibitem [{\citenamefont {Ruiz}\ \emph {et~al.}(2018)\citenamefont {Ruiz},
  \citenamefont {Shapiro},\ and\ \citenamefont {Tsokaros}}]{Ruiz2018}%
  \BibitemOpen
  \bibfield  {author} {\bibinfo {author} {\bibfnamefont {M.}~\bibnamefont
  {Ruiz}}, \bibinfo {author} {\bibfnamefont {S.~L.}\ \bibnamefont {Shapiro}}, \
  and\ \bibinfo {author} {\bibfnamefont {A.}~\bibnamefont {Tsokaros}},\ }\href
  {\doibase 10.1103/PhysRevD.97.021501} {\bibfield  {journal} {\bibinfo
  {journal} {Phys. Rev. D}\ }\textbf {\bibinfo {volume} {97}},\ \bibinfo
  {pages} {021501} (\bibinfo {year} {2018})}\BibitemShut {NoStop}%
\bibitem [{\citenamefont {Shibata}\ \emph {et~al.}(2017)\citenamefont
  {Shibata}, \citenamefont {Fujibayashi}, \citenamefont {Hotokezaka},
  \citenamefont {Kiuchi}, \citenamefont {Kyutoku}, \citenamefont {Sekiguchi},\
  and\ \citenamefont {Tanaka}}]{Shibata2017}%
  \BibitemOpen
  \bibfield  {author} {\bibinfo {author} {\bibfnamefont {M.}~\bibnamefont
  {Shibata}}, \bibinfo {author} {\bibfnamefont {S.}~\bibnamefont
  {Fujibayashi}}, \bibinfo {author} {\bibfnamefont {K.}~\bibnamefont
  {Hotokezaka}}, \bibinfo {author} {\bibfnamefont {K.}~\bibnamefont {Kiuchi}},
  \bibinfo {author} {\bibfnamefont {K.}~\bibnamefont {Kyutoku}}, \bibinfo
  {author} {\bibfnamefont {Y.}~\bibnamefont {Sekiguchi}}, \ and\ \bibinfo
  {author} {\bibfnamefont {M.}~\bibnamefont {Tanaka}},\ }\href {\doibase
  10.1103/PhysRevD.96.123012} {\bibfield  {journal} {\bibinfo  {journal} {Phys.
  Rev. D}\ }\textbf {\bibinfo {volume} {96}},\ \bibinfo {pages} {123012}
  (\bibinfo {year} {2017})}\BibitemShut {NoStop}%
\bibitem [{\citenamefont {Alsing}\ \emph {et~al.}(2018)\citenamefont {Alsing},
  \citenamefont {Silva},\ and\ \citenamefont {Berti}}]{Alsing2018}%
  \BibitemOpen
  \bibfield  {author} {\bibinfo {author} {\bibfnamefont {J.}~\bibnamefont
  {Alsing}}, \bibinfo {author} {\bibfnamefont {H.~O.}\ \bibnamefont {Silva}}, \
  and\ \bibinfo {author} {\bibfnamefont {E.}~\bibnamefont {Berti}},\ }\href
  {\doibase 10.1093/mnras/sty1065} {\bibfield  {journal} {\bibinfo  {journal}
  {Monthly Notices of the Royal Astronomical Society}\ }\textbf {\bibinfo
  {volume} {478}},\ \bibinfo {pages} {1377} (\bibinfo {year}
  {2018})}\BibitemShut {NoStop}%
\bibitem [{\citenamefont {Bedaque}\ and\ \citenamefont
  {Steiner}(2015)}]{Bedaque2015}%
  \BibitemOpen
  \bibfield  {author} {\bibinfo {author} {\bibfnamefont {P.}~\bibnamefont
  {Bedaque}}\ and\ \bibinfo {author} {\bibfnamefont {A.~W.}\ \bibnamefont
  {Steiner}},\ }\href {\doibase 10.1103/PhysRevLett.114.031103} {\bibfield
  {journal} {\bibinfo  {journal} {Phys. Rev. Lett.}\ }\textbf {\bibinfo
  {volume} {114}},\ \bibinfo {pages} {031103} (\bibinfo {year}
  {2015})}\BibitemShut {NoStop}%
\end{thebibliography}%

\end{document}